\documentclass[12pt]{iopart}

\usepackage{iopams}  
\usepackage{graphicx}
\usepackage{aas_macros} 

\begin{document}

\title[Cherenkov Telescopes]{Ground based gamma-ray astronomy with Cherenkov Telescopes}

\author{Jim Hinton}

\address{School of Physics \& Astronomy, University of Leeds, Leeds LS2 9JT, UK}

\ead{j.a.hinton@leeds.ac.uk}

\begin{abstract}

Very-high-energy ($>$100 GeV) $\gamma$-ray astronomy is emerging as
an important discipline in both high energy astrophysics and
astro-particle physics. This field is currently dominated by Imaging Atmospheric-Cherenkov Telescopes
(IACTs) and arrays of these telescopes. Such arrays have achieved 
the best angular resolution and energy flux sensitivity in the $\gamma$-ray
domain and are still far from the fundamental limits of the technique.
Here I will summarise some key aspects of this technique
and go on to review the current status of the major instruments and
to highlight selected recent results.

\end{abstract}

\maketitle

\section{Introduction}

Ground-based $\gamma$-ray astronomy effectively began in 1989, with
the first detection of a TeV $\gamma$-ray source, the Crab Nebula,
with the 10~m Cherenkov telescope of the Whipple
Observatory~\cite{Whipple:crab}.  Eighteen years on, Cherenkov
telescopes have been used to detect 68 very high energy (VHE;
$E\,>\,100$~GeV) sources, firmly establishing a new astronomical
domain. The key advantage of ground based instrumentation over
satellite-based GeV instruments such as EGRET and the upcoming GLAST
large area telescope (LAT) is collection area. The typical effective
collection area of single Cherenkov telescope is $10^{5}$ m$^2$,
almost five orders of magnitude larger than can realistically be
achieved via direct detection in space.  The major advantage of the
Imaging Atmospheric-Cherenkov Telescope (IACT) technique with respect to other ground-based approaches (which
are described elsewhere in this volume) is the precision with which 
the properties of the primary  $\gamma$-ray can be reconstructed.
The angular resolution achievable is currently limited only by the number of
Cherenkov photons collected, with the theoretical
limit close to 30$''$ at 1 TeV, an order of magnitude better than
current instruments have achieved~\cite{Hofmann:Performance}.
Indeed, the typical angular resolution of $0.1^{\circ}$, whilst not
impressive in comparison to that achieved over much of the 
electromagnetic spectrum, is the best at any energy above 
$\sim$0.1~MeV. 

Here I will summarise the Cherenkov Telescope technique and then
summarise the results produced using these instruments.

\section{The Imaging Atmospheric-Cherenkov Technique}

\subsection{Cherenkov light from air-Showers}

\begin{figure}
\begin{center}
\includegraphics[width=0.49\textwidth,height=0.59\textwidth]{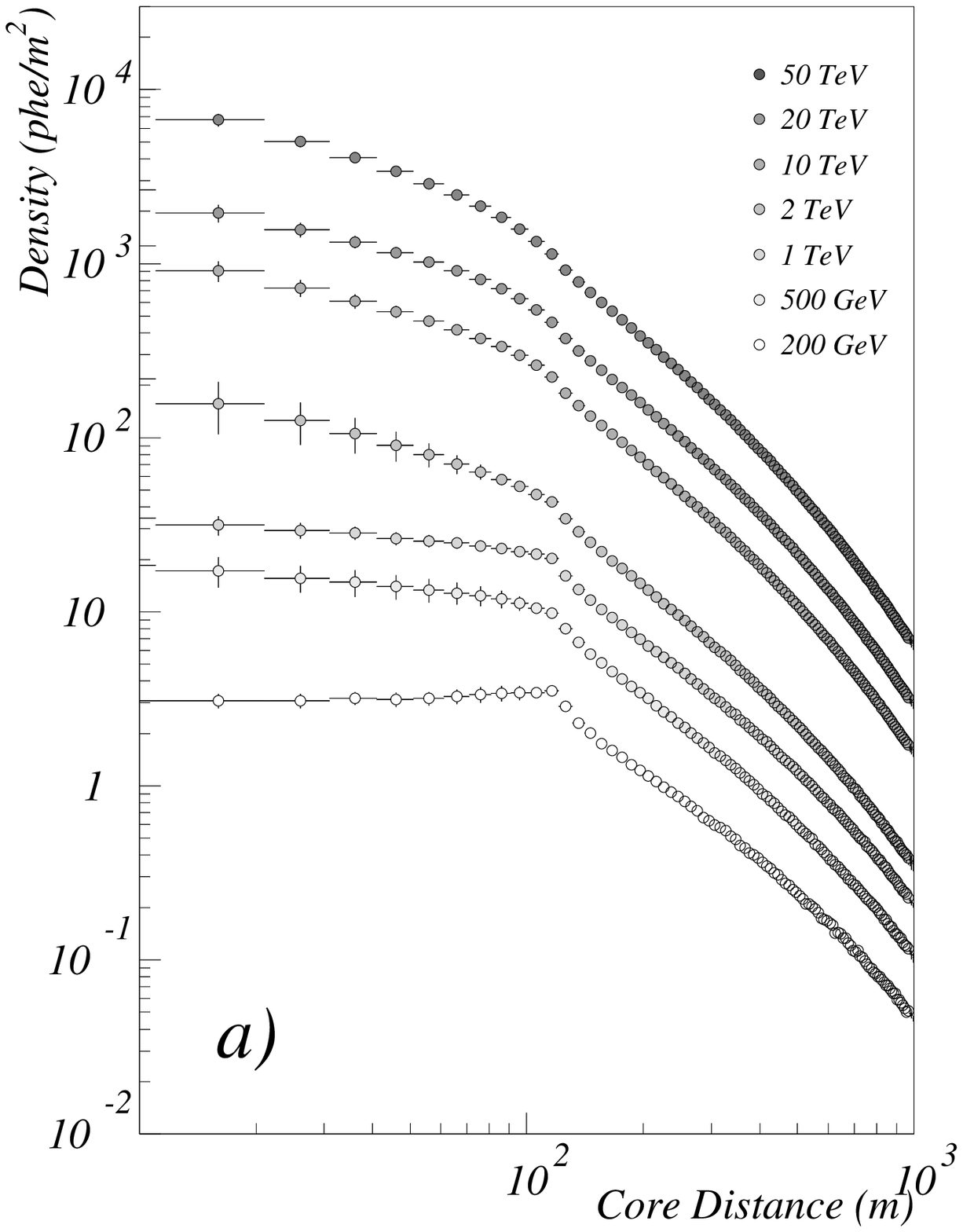}\includegraphics[width=0.49\textwidth,height=0.59\textwidth]{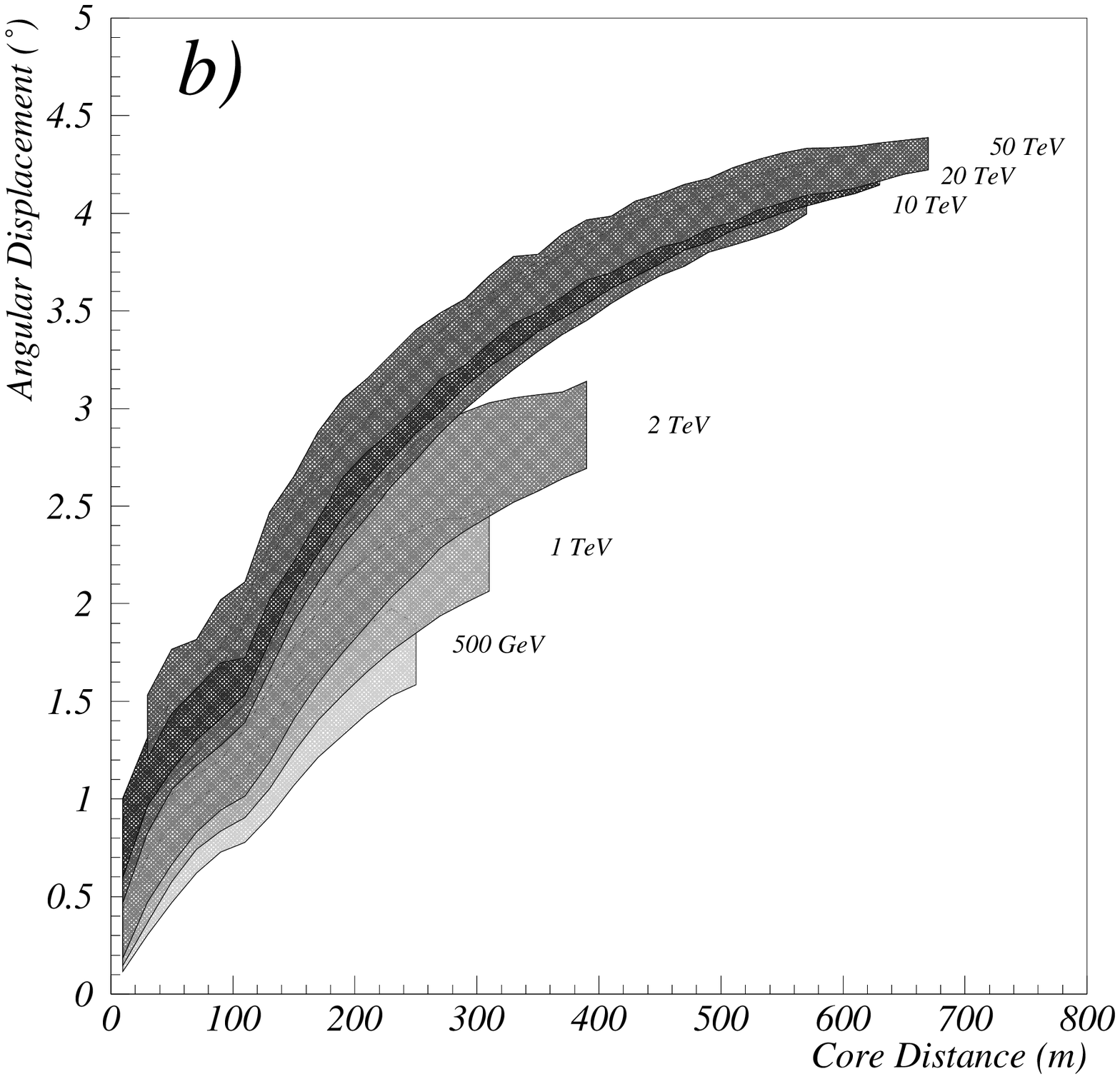}
\caption{a) The lateral distribution of Cherenkov light for $\gamma$-ray primaries of various energies at 2400~m altitude.
b) The displacement of the centroid of the Cherenkov image as a function of impact distance. Both plots are reproduced from \cite{Biller:WideFoV}.}
\label{fig:ldf}
\end{center}
\end{figure}

High energy ($>$ GeV) photons entering the Earth's atmosphere initiate
electromagnetic cascades, via the processes of electron
pair-production and subsequent bremsstrahlung. The number of electrons
at the point of maximum development of the cascade is closely
proportional to the primary energy and the atmospheric depth of this
maximum increases logarithmically with energy. For a 1 TeV
photon-initiated air-Shower this maximum occurs at a depth of
$\sim$300 g cm$^{-2}$, or at $\sim$10~km above sea level (a.s.l.) for
a vertically incident photon. Electrons and positrons in the shower
with energies greater than $m_{e}c^{2}/\sqrt{1-n^{-2}}$ will emit
Cherenkov light. This threshold corresponds to $\approx$20~MeV in air
at sea level and roughly twice that at 10~km a.s.l. The yield of
Cherenkov light is proportional to the total track length of all
particles (in the ultra-relativistic limit) which is in turn
proportional to the primary energy. In this way an image of the
cascade in Cherenkov light provides a pseudo-calorimetric measurement
of the shower energy. The opening angle of Cherenkov light in air is
roughly 1$^{\circ}$ and hence the photons produced around shower
maximum arrive at typical observation heights of $\sim$2000~m
a.s.l. in a `light-pool' of $\sim$120~m radius. As can be seen from
figure~\ref{fig:ldf}, the density of photons within this light pool is
roughly 100 photons (of wavelength 300--600~nm) per square-metre per TeV
of primary energy.  For a typical instrumental efficiency of 10\%
(reflectivity of mirror surfaces; quantum efficiency of
photo-sensors), primary reflectors of $\sim$100 m$^{2}$ area are
required to produce images containing 100 photoelectrons for 100~GeV
$\gamma$-ray showers.  The fall-off in density outside this region is
rather rapid, but as figure~\ref{fig:ldf} illustrates, at very high
energies ($>10$ TeV), distant showers (with impact distances of
several hundred metres) are visible even with modest sized
telescopes. The `flash' of Cherenkov light at the ground lasts only a
few nanoseconds. Fast photo-sensors and electronics are therefore
employed to resolve the faint Cherenkov signal against the night sky
background (NSB) light, which has a typical rate of 1 photon-electron
$\times (\Delta t / 10 \rm{ns})(\theta/ 0.1^{\circ})^{2} (A / 100
\rm{m}^{2})$. Outside periods of astronomical darkness the background
light level is much higher, imposing a $\sim$10\% duty cycle on IACT
observations (at least for low energy threshold measurements, see
\cite{MAGIC:moon}).

As most Cherenkov light from TeV showers is produced around the point
of maximum development of the shower, the intensity of light at ground level scales approximately
as $1/d^{2}$ where $d$ is the distance to the point of shower maximum,
and conversely the area covered by the Cherenkov light pool is proportional
to $d^{2}$. Observations at lower altitudes or equivalently greater zenith angles
can be used to increase collection area, at the expense of a higher energy threshold
(e.g. \cite{GC:Whipple04, HESS:mrk421}).
Conversely, very high altitude observatories have been suggested as a natural way 
to achieve lower energy thresholds (e.g. \cite{5at5}). 


Images of $\gamma$-ray showers have typical rms widths and lengths of
$\sim$0.1$^{\circ}$ and $\sim$0.3$^{\circ}$, respectively (at
$\sim$1~TeV, 100~m impact distance). Pixelisation of not much greater
than 0.1$^{\circ}$ is therefore required to resolve the showers. The
displacement of the shower image centroid is directly related to the
impact distance of the shower axis with respect to the telescope (see
figure~\ref{fig:ldf}b). At high energies, the effective collection
area for a Cherenkov telescope is therefore often limited by the size
of the field of view (FoV), rather than by mirror area (or
equivalently photon density).  Figure~\ref{fig:ldf}b demonstrates that
a FoV of at least $\sim4^{\circ}$ is desirable for TeV
observations. In fact it is difficult to provide a field of view much
bigger than this and maintain an optical point-spread-function (PSF)
on the scale of $<$1 pixel, without the use of secondary optics and
hence additional cost and complexity. Cost has also so far limited the
total pixel number to less than $\sim$1000, again with implications
for the field of view achievable.

\subsection{$\gamma$/Hadron separation}

As discussed above a rather simple light collector of area 100~m$^{2}$
is sufficient to detect $\sim$100~GeV $\gamma$-ray showers if placed at
mountain altitudes. The major challenge of the Cherenkov technique is
the presence of an, at one time, overwhelming background of air-showers
initiated by cosmic ray protons and nuclei. For example, for
observations with the H.E.S.S. telescope array, the rate of detected
photons from the brightest steady sources is still only $\sim$0.1\% of
the rate of background showers.  Fortunately, showers initiated by TeV
protons and nuclei differ in many respects from $\gamma$-ray
showers. Much of the energy in the primary is transferred to pions
produced in the first few interactions. The neutral pions decay to
produce electromagnetic \emph{sub-showers}, with the charged pions
decaying to produce muons.  Single muons reaching ground level produce
ring images if impacting the telescope dish, or arcs at larger impact
distances.  The sub-showers often result in substructure in images and
the larger transverse angular momentum in hadronic interactions leads
to showers that are generally wider than those of $\gamma$-rays.  For a
given primary energy, hadronic showers also produce less Cherenkov
light (a factor $\sim$2--3 at TeV energies), due to the energy 
channelled into neutrinos and into high energy muons and hadrons in the
shower core.

The primary discriminator between hadron and $\gamma$-ray initiated
showers is therefore the width of the Cherenkov image. The
breakthrough in the technique was the recognition by Hillas in the
1980's that the measurement and simple parameterisation of images
allows very effective background rejection~\cite{hillas85:technique,
hillas96:technique}.  Several more sophisticated background rejection
and shower reconstruction methods have now been developed (see for
example \cite{CAT:model}, \cite{deNaurois:Model},
\cite{HESS:model_marianne}) but the ``Hillas parameter'' approach
remains the standard in the field.

Despite the rejection of the vast majority of the background using image
cuts, the correct modelling and subtraction of the remaining background is
a major challenge and a potential sources of systematic errors, see
\cite{Berge:background} for a recent summary.

\subsection{Stereoscopic Measurements}

The desirability of multiple telescope observations of individual
air-showers was first demonstrated by the HEGRA collaboration
\cite{HEGRA:performance1}.  The first advantage occurs at the trigger
level: an array with a multi-telescope trigger system removes the vast
majority of single muons and also many hadron initiated showers.
For a dead-time limited system this may allow a lower trigger (and
hence energy) threshold, see for example \cite{HESS:trigger}. Other
advantages arise at the analysis stage, primarily in the
reconstruction of the shower geometry and hence in the reconstruction
of the direction and energy of the primary $\gamma$-ray. Shower axis
reconstruction with a single Cherenkov telescope is possible using the
length of the image to estimate the angular distance to the source
position \cite{mono_rec}. However, the multiple views of the shower
provided by stereoscopic observations allow a more accurate
determination of the shower direction using
the intersection of the directions defined by the major axes of the
images recorded in each camera.  In a similar way, the shower core
location can be better established, leading to improved energy
resolution (due the dependence of Cherenkov light intensity on impact
distance, see figure~\ref{fig:ldf}).  The improved shower geometry
also leads to better hadron rejection, the primary rejection parameter
\emph{width} can be replaced by \emph{mean scaled width}, normalising
based on expectations for $\gamma$-ray showers (for a given image
amplitude and impact distance) and averaged over all telescopes (see
for example \cite{HEGRA:performance2}).  The optimal separation of
telescopes in an array seems to be close to the radius of the
Cherenkov light-pool ($\sim$100~m), with closer spacing improving
low-energy performance at the expense of effective collection area at
higher energies (and vice versa).

\subsection{The use of timing information}

The arrival time structure of Cherenkov images provides a potential additional discriminator
against the hadronic background. Single muons certainly exhibit a characteristic very fast
time profile~\cite{Mirzoyan:Muons} and timing information may be useful in rejecting 
hadronic showers (see for example \cite{Cherenkov:Timing}).
Several instruments of the current generation record digitised waveforms
and there is evidence to suggest that a significant improvement in 
background rejection, and hence sensitivity, can be achieved using this
information --- at least for single telescopes \cite{MAGIC:timing}. 
However, these studies are still at a relatively early stage and it is not 
yet clear if substantial gains are possible for stereoscopic systems.

\section{Current Instruments}

\begin{figure}
\begin{center}
\includegraphics[width=1.02\textwidth]{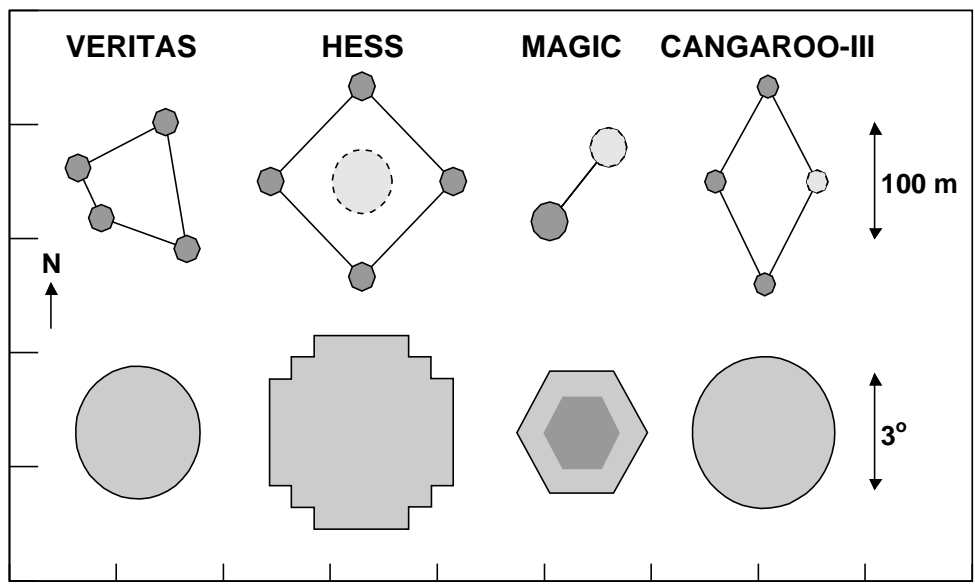}
\caption{Comparison of the array layout (top) and camera field of view (bottom)
for the major 
atmospheric Cherenkov detectors. The radius of the circles representing the 
telescope dishes has been doubled for clarity. The dashed circles indicate
telescopes currently under construction. The darker region at the centre of
the MAGIC camera illustrates the region with smaller pixel size.}
\label{fig:arrays}
\end{center}
\end{figure}

Table~\ref{tab:inst} summarises the characteristics of currently
operating Imaging Cherenkov telescopes and arrays, together with
some important decommissioned systems. In the following I will discuss
the current systems, with emphasis on the major Cherenkov telescope systems
illustrated in Figure~\ref{fig:arrays}. 

\begin{table}
\footnotesize\rm
\caption{Principle characteristics of currently operating (and
selected historical) IACTs and IACT arrays. The
energy threshold given is the approximate trigger-level (rather than
post-analysis) threshold for observations close to zenith. The
approximate sensitivity is expressed as the minimum flux (as a
percentage of that of the Crab Nebula: $\approx 2\times10^{-11}$
photons cm$^{-2}$ s$^{-1}$ above 1 TeV) of a point-like source
detectable at the $5\sigma$ significance level in a 50~hour
observation. In the cases where this number has not been provided by
experimental collaborations, it is estimated here from published
detections.  $^{\star}$ No refereed publications from this instrument
exist and it's sensitivity is therefore very difficult to
estimate. $^{\dagger}$ These instruments have pixels of two different
sizes.  }
\begin{tabular}{|l|c|c|c|c|c|c|c|c|c|c|} \hline
Instrument & Lat. & Long. & Alt. & Tels. & Tel. Area & Total A. & Pixels & FoV & Thresh. & Sensitivity \\
           & ($^{\circ}$) & ($^{\circ}$) & (m) &            & (m$^{2})$  & (m$^{2}$)  &            & ($^{\circ}$) & (TeV) & (\% Crab) \\\hline 
H.E.S.S. & -23 & 16 & 1800 & 4 & 107 & 428 & 960 & 5 & 0.1 & 0.7 \\ 
VERITAS & 32 & -111 & 1275 & 4 & 106 & 424 & 499 & 3.5 & 0.1 & 1 \\ 
MAGIC & 29 & 18 & 2225 & 1 & 234 & 234 & 574 &  3.5$^{\dagger}$ & 0.06 & 2 \\ 
CANGAROO-III & -31 & 137 & 160 & 3 & 57.3 & 172 & 427 & 4 & 0.4 & 15 \\
Whipple & 32 & -111 & 2300 & 1 & 75 & 75 & 379 & 2.3 & 0.3 & 15 \\
Shalon & 43 & 77 & 3338 & 1 & 11.2 & 11.2 & 144 & 8 & 0.8 & $^{\star}$ \\
TACTIC & 25 & 78 & 1300 & 1 & 9.5 & 9.5 & 349 & 3.4 & 1.2 & 70 \\\hline
\emph{HEGRA} & 29 & 18 & 2200 & 5 & 8.5 & 43 & 271 & 4.3 & 0.5 & 5 \\
\emph{CAT} & 42 & 2 & 1650 & 1 & 17.8 & 17.8 & 600 & 4.8$^{\dagger}$ & 0.25 & 15 \\
\hline
\end{tabular}
\label{tab:inst}
\end{table}

\emph{The High Energy Stereoscopic System} (H.E.S.S.) is an array of 4
Cherenkov telescopes situated in the Khomas highlands of
Namibia~\cite{HESS:status}. Completed in early 2004, H.E.S.S. was the
first of the new generation of Cherenkov Telescope arrays to become
fully operational.  The H.E.S.S. telescopes have 13~m diameter
(107~m$^2$) dishes and a focal length of 15~m~\cite{HESS:optics}.  The
Davis-Cotton optical design allows a wide field with reasonable
off-axis optical PSF.  The optical PSF has an 80\% containment radius
of $1.4'$ on-axis, with the diameter becoming comparable to the camera
pixel size ($0.16^{\circ}$) only at the edge of the field-of-view.
The cameras consist of 960 photomultiplier tube pixels, with signal
acquisition via 1~GHz analogue ring samplers (ARSs), in normal
operation however only the 16~ns integrated signal is read out to
reduce dead-time. H.E.S.S. utilises an array level trigger which for
normal operations requires a telescope multiplicity of
two~\cite{HESS:trigger}.  Construction has just begun on an upgrade to
H.E.S.S. consisting of a single very large (600~m$^{2}$) parabolic
telescope at the centre of the phase-1 array~\cite{HESS:phaseII}

\emph{The MAGIC telescope} on La Palma, in the Canary Islands, is, at
17~m diameter, the largest single Cherenkov telescope in operation and
also reaches the lowest trigger-level energy threshold (at $\approx$60
GeV) \cite{MAGIC:status}.  The design of the instrument was driven by
two goals, to be able to rapidly slew the telescope at respond to
$\gamma$-ray burst (GRB) alerts, and to achieve an energy threshold as
low as possible given the size of the dish. The light-weight
construction allows a slewing speed of $\sim$5$^{\circ}$/s ($\sim$3
times faster than the H.E.S.S. telescopes)~\cite{MAGIC:technical}.  A
parabolic dish removes the time dispersion due to optical path length
differences and a recent upgrade to 2~GHz waveform sampling allows
exploitation of the timing information in the Cherenkov front (see
above).  The advantages of a stereoscopic system (as discussed above)
motivated the second phase of the MAGIC project: the construction of a
second 17~m telescope 85~m from the first. This second telescope is
currently under construction and will have a camera with uniform pixel
size and an increased trigger region~\cite{MAGIC-II:camera}.

\emph{VERITAS} (The Very Energetic Radiation Imaging Telescope Array
System) is an array of four 12~m diameter telescopes situated at the
base-camp of the Whipple Observatory in
Arizona~\cite{VERITAS:first_tel,VERITAS:status}.  First light for the
full four telescope array occurred early in 2007.  The overall design,
and hence the sensitivity, of the array is rather similar to that of
H.E.S.S. \cite{VERITAS:icrc}. VERITAS has the advantage of 500~MHz
flash ADCs, but a somewhat narrower field-of-view, which may limit
performance at high energies.

The \emph{CANGAROO-III} instrument \cite{CANGAROO3:status} is a four
telescope system continuing the CANGAROO project
\cite{CANGAROO:3.8m,CANGAROO2:status} on a site near Woomera,
Australia.  The array of 3 new 10~m diameter telescopes was completed
in 2004. A fourth telescope, that of the CANGAROO-II instrument is
currently not included in the array trigger, pending a camera upgrade.
Some controversy surrounded some detections using CANGAROO and
CANGAROO-II, but all disagreements with results from the
H.E.S.S. array have now been resolved by observations with the more
sensitive CANGAROO-III instrument~\cite{CANGAROO:icrc}.

Other currently operating Cherenkov instruments include the TACTIC
imaging telescope~\cite{TACTIC:NIM} and the non-imaging PACT
system~(see e.g. \cite{PACT:agn}), both located in India. PACT is to my knowledge the
only remaining non-imaging Cherenkov detector. In the recent past,
several converted solar power stations were used to make Cherenkov
observations at rather low energies (50--150 GeV thresholds) but
with modest hadron rejection power and hence sensitivity.  These
instruments were operated by the CELESTE~\cite{CELESTE:status},
STACEE~\cite{STACEE:status}, Solar Two (later CACTUS)~\cite{SolarTwo:status}, 
and GRAAL~\cite{GRAAL:crab} collaborations.

\section{Recent Science Highlights} 

\subsection{Galactic Sources}

\begin{figure}
\begin{center}
\includegraphics[width=0.99\textwidth]{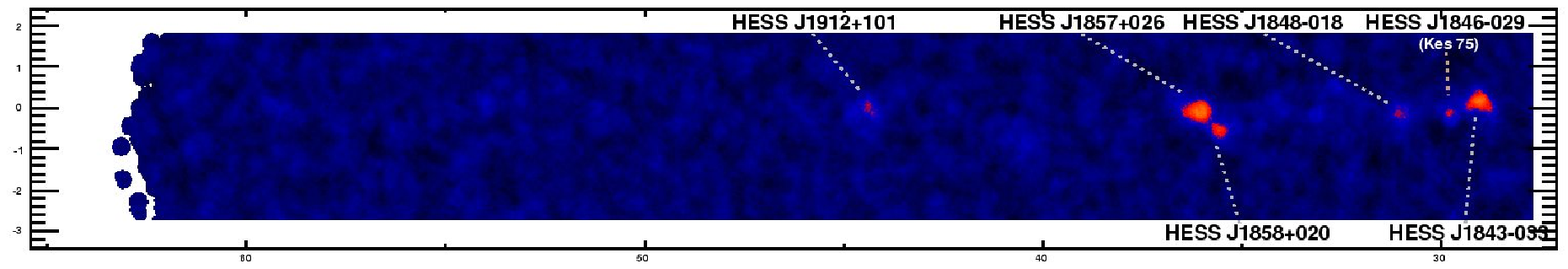}
\includegraphics[width=0.99\textwidth]{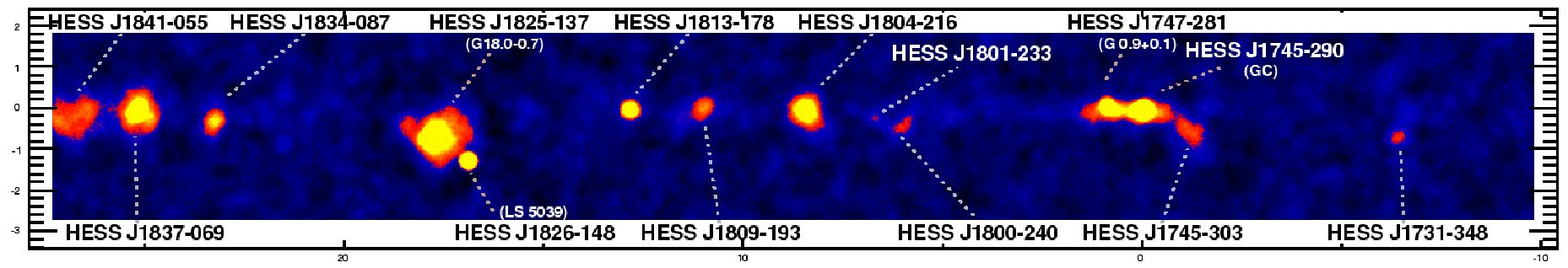}
\includegraphics[width=0.99\textwidth]{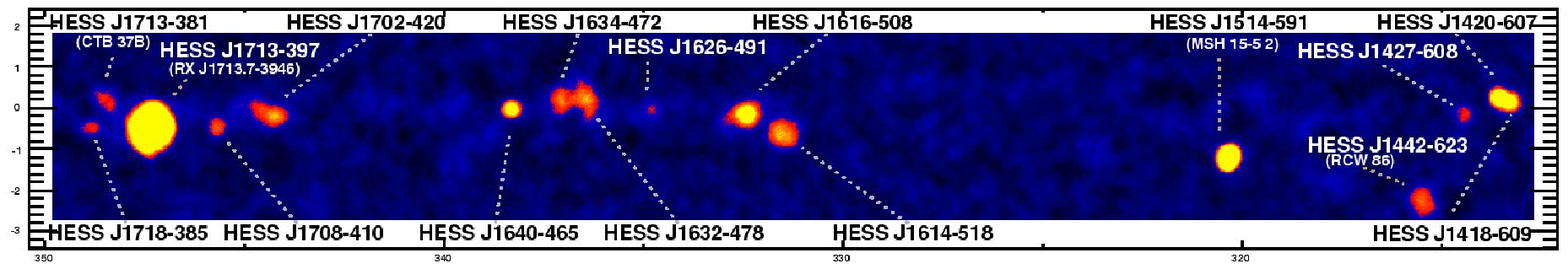}
\includegraphics[width=0.99\textwidth]{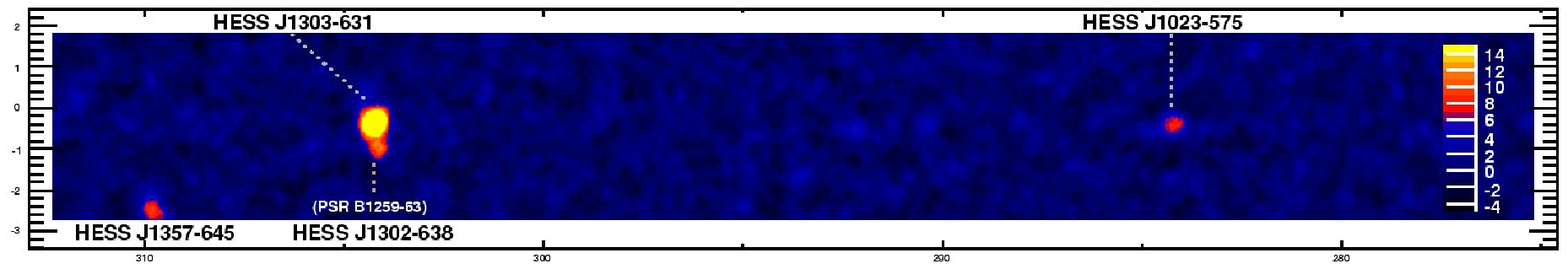}
\caption{The H.E.S.S. survey of the inner galaxy in $\sim$1~TeV $\gamma$-rays.
The colour scale shows the statistical significance for an excess within an
$0.22^{\circ}$ radius at each position. Image courtesy of the H.E.S.S. Collaboration.}
\label{fig:scan}
\end{center}
\end{figure}

The recent one order of magnitude growth in the number of known
\emph{galactic} VHE $\gamma$-ray sources is largely a consequence of
the survey of the galactic plane conducted by the H.E.S.S.
collaboration between 2004 and
2007~\cite{HESS:scanpaper1,HESS:scanpaper2, HESS:scanicrc}.
Figure~\ref{fig:scan} shows the current extent of this scan, which now
covers essentially the whole inner galaxy: $-85^{\circ} < l <
60^{\circ}, -2.5^{\circ} < b < 2.5^{\circ}$ \cite{HESS:scanicrc}.
Most of the 52 currently known galactic TeV sources remain
unidentified. This is in part due to the difficulty of identifying
extended sources with no clear sub-structure.  Nonetheless, several
methods of identification have been successfully applied and the
situation is much more favourable than that in the GeV band where only
one galactic source class (pulsars) has been unambiguously identified.

Table~\ref{tab:gal} lists those galactic VHE sources for which a multi-wavelength 
counterpart can be considered to be well established (note that such a classification
is subjective and this is simply my personal view). There are three classes of such
objects: supernova remnants, pulsar wind nebulae and binary systems: 

\begin{table}
  \footnotesize\rm
  \caption{ Galactic very high energy $\gamma$-ray sources with well
    established multi-wavelength counterparts.  The instrument used to
    discover the VHE emission is given together with the year of
    discovery.  Fluxes are approximate, and expressed as a percentage
    of the flux from the Crab Nebula in the TeV range, $^{\star}$
    indicates variable emission.  The photon index $\Gamma$
    ($dN/dE\,\propto\,E^{-\Gamma}$) is given together with its
    statistical error, systematic errors are typically $\sim$0.2.  The
    final column gives citations to the respective discovery paper and
    other papers marking significant steps towards the identification of
    the source. These associations were established through a
    range of methods, which are given in the table in abbreviated
    form: \emph{Pos}: The position of the centroid of the VHE emission
    can be established with sufficient precision that there is no
    ambiguity as to the low energy counterpart. In practise this is
    usually only possible for point-like sources. \emph{Mor}: There is
    a match between the $\gamma$-ray morphology and that seen at other
    (usually X-ray) wavelengths. This requires sources extended well
    beyond the typical angular resolution of IACTs
    ($\sim$0.1$^{\circ}$). \emph{EDMor}: Energy-dependent morphology
    which approaches the position/morphology seen at other wavelengths
    at some limit, and is consistent with our physical understanding
    of the source. \emph{Var}: $\gamma$-ray variability correlated
    with that in other wavebands. \emph{Per}: periodicity in the
    $\gamma$-ray emission matching that seen at other
    wavelengths. Note that all these objects are X-ray sources.}
    \begin{tabular}{|l|l|l|l|l|l|l|l|l|} \hline 
Object	&  Discovered     &Year	&  Type	      & Method	& Flux  & Index	&Ref.  \\\hline
PSR B1259$-$63
                &  HESS	  & 2005 &  Binary  &Pos/Var & 7$^{\star}$	& $2.7\pm0.2$	& \cite{HESS:psrb1259}   \\
LS\,5039	&  HESS	  & 2005 &  Binary     &Pos/Per	& 3$^{\star}$	& $2.12\pm0.15^{\star}$	& \cite{HESS:ls5039p1,HESS:ls5039p2}  \\
LS\,I\,+61\,303  &  MAGIC  & 2006 &  Binary     &Pos/Var	& 16$^{\star}$	& $2.6\pm0.2$	&  \cite{MAGIC:lsi61,VERITAS:lsi61}   \\
RX\,J1713.7$-$3946& CANGAROO & 2000 &  SNR Shell  &	Mor	& 66	& $2.04\pm0.04^{\dagger}$ &  \cite{CANGAROO:rxj1713,HESS:rxj1713p1,HESS:rxj1713p2} \\
Vela Junior	&  CANGAROO & 2005	&  SNR Shell  &	Mor	& 100	& $2.24\pm0.04$	& \cite{CANGAROO:velajnr,HESS:velajnr,HESS:velajnr2}  \\
RCW\,86		&  HESS	  & 2007 &  SNR Shell  &	Mor	& 5-10?	& -	& \cite{HESS:rcw86_icrc} \\
Cassiopeia\,A	&  HEGRA  & 2001 &  SNR	      &	Pos	& 3	&$2.5\pm0.4$	& \cite{HEGRA:casA}    \\
Crab Nebula	&  Whipple& 1989 &  PWN	      &	Pos	& 100	& $2.49\pm0.06$	& \cite{Whipple:crab,Whipple:crabspec} \\
MSH\,15-52	&  HESS	  & 2005 &  PWN	      &	Mor	& 15	& $2.27\pm0.03$	& \cite{HESS:msh1552}  \\
Vela\,X		&  HESS	  & 2006 &  PWN	      &	Mor	& 75	& $1.45\pm0.09^{\dagger}$ & \cite{HESS:velax}    \\
HESS\,J1825$-$137
                &  HESS	  & 2005 &  PWN	      & EDMor	& 12    & $2.26\pm0.03^{\dagger}$      & \cite{HESS:scanpaper1,HESS:1825p1,HESS:1825p2}  \\
PSR\,J1420$-$6049
                &  HESS	  &2006&  PWN 	      &	Mor	&  7	& $2.17\pm0.06$ & \cite{HESS:kookaburra}  \\
The Rabbit	&  HESS	  &2006&  PWN	      &	Mor	&  6	& $2.22\pm0.08$ & \cite{HESS:kookaburra} \\
G\,0.9+0.1	&  HESS	  &2005&  PWN	      &	Pos     &  2    & $2.40\pm0.11$ & \cite{HESS:g09}\\
\hline
\end{tabular}
\label{tab:gal}
\end{table}

\begin{figure}
\begin{center}
\includegraphics[width=0.99\textwidth]{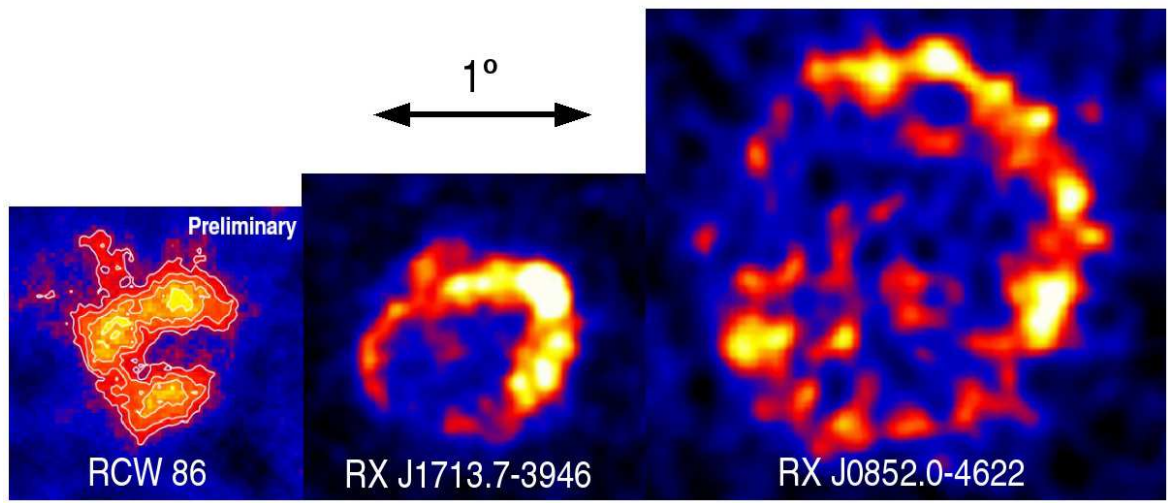}
\caption{The shell-type TeV $\gamma$-ray supernova remnants: 
RCW\,86 \cite{HESS:rcw86_icrc}, 
RX\,J1713.7$-$3946 \cite{HESS:rxj1713p3}
and RX\,J0852.0$-$4622 (\emph{Vela Junior}) \cite{HESS:velajnr2}. 
All images are smoothed and were obtained using H.E.S.S.}
\label{fig:snrs}
\end{center}
\end{figure}

\emph{Shell-type Supernova Remnants} have long been considered as the likely 
acceleration site for the bulk of the galactic cosmic rays. As such they 
were prime targets for the first Cherenkov telescopes. The first (subsequently
confirmed) SNRs to be detected at TeV energies were RX\,J1713.7$-$3946~\cite{CANGAROO:rxj1713}  
and Cassiopeia\,A~\cite{HEGRA:casA}, by the CANGAROO and HEGRA collaborations, respectively.
The emission from RX\,J1713.7$-$3946 was resolved using H.E.S.S. as a shell 
with very similar morphology to that seen in non-thermal X-rays~\cite{HESS:rxj1713p1}.
There are now three such resolved TeV shell SNR, shown in figure~\ref{fig:snrs}.
These images demonstrate the existence of $>$ TeV particles in the expanding shocks
of these objects. However, the nature of the particles dominantly responsible for the 
TeV emission is still hotly debated.

\begin{figure}
\begin{center}
\includegraphics[width=0.9\textwidth]{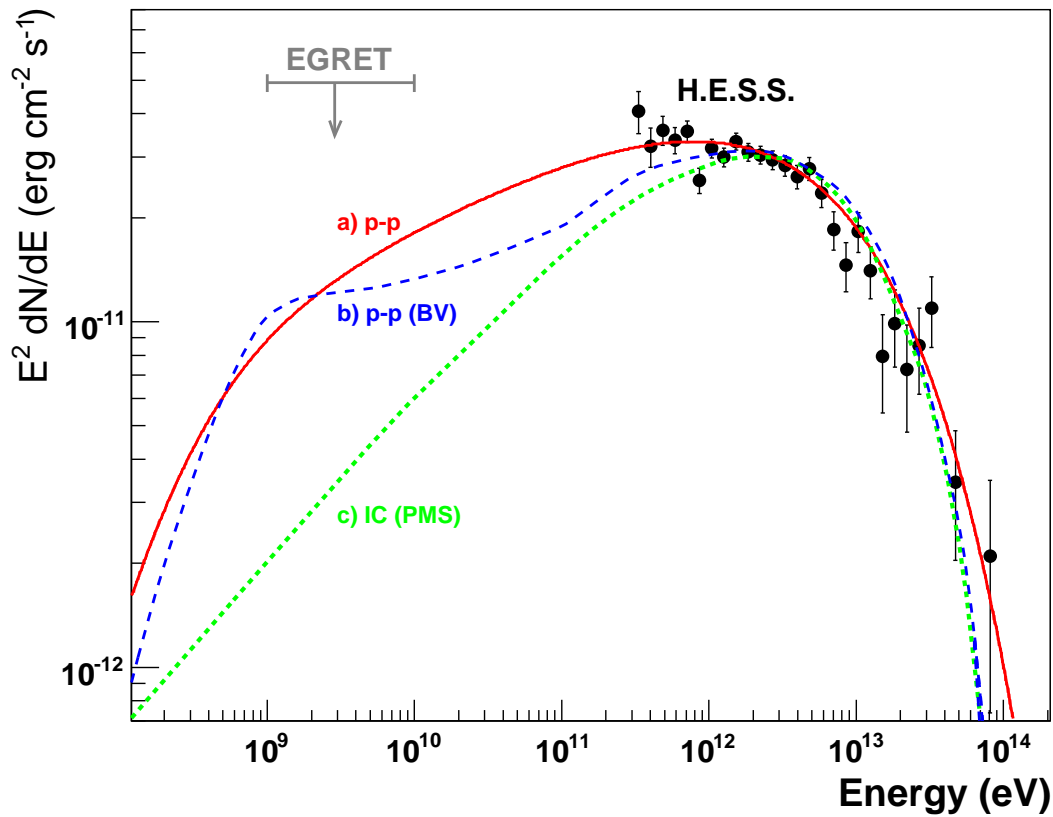}
\caption{The spectral energy distribution of RX\,J1713.7$-$3946
in the $\gamma$-ray range. The TeV data points are taken from \cite{HESS:rxj1713p3},
model curves are given for three scenarios: 
(a) a fit of the function $dN/dE\,\propto\,E^{-\Gamma}\exp{-\sqrt{E/E_{0}}}$, 
the approximate form expected for interacting protons with a energy distribution 
following a power-law with exponential cut-off, see \cite{HESS:rxj1713p3} and \cite{Kelner:pp}.
(b) hadronic emission as calculated by \cite{Berezhko:rxj1713}, and
(c) inverse Compton emission as calculated by \cite{Porter:ic_snr}.
}
\label{fig:1713spec}
\end{center}
\end{figure}

Two basic scenarios have been widely discussed for the best measured 
object RX\,J1713.7$-$3946: 1) the $\gamma$-ray signal is inverse Compton
emission from the same population of high energy electrons responsible
for the synchrotron X-ray emission; 2) hadronic interactions of protons
and nuclei lead to $\gamma$-ray emission via the decay of neutral pions.
The first case is supported by correlation between X-ray and TeV emission,
but implies a magnetic field close to 10 $\mu$G, uncomfortably low in 
many models. The spectral shape of the $\gamma$-ray emission presents one
way to break this ambiguity. Figure~\ref{fig:1713spec} compares the
measured spectral energy distribution of RX\,J1713.7$-$3946 to 
expectations for three different scenarios. At present, hadronic models
(see for example~\cite{Berezhko:rxj1713}) seem favoured for this object.

Another method to identify $\gamma$-ray emission as hadronic in origin
is to establish a spatial correlation of the $\gamma$-ray emission
with available target material. Indeed, such correlations seem to be
present for the two (somewhat older: $\sim$10$^{4-5}$ years cf
$\sim$1000 years) SNRs W\,28 and IC\,443 in
H.E.S.S.~\cite{HESS:w28_icrc}, MAGIC~\cite{MAGIC:ic443} and
VERITAS~\cite{VERITAS:ic443_icrc} data, strongly suggesting that these
objects are accelerating hadronic cosmic-rays. Such a correlation is
also seen in the region of the Galactic Centre. However, in that
particular case the acceleration site of the cosmic rays is not
clear~\cite{HESS:gc_diffuse}.

\emph{Pulsar Wind Nebulae} (PWN) have now emerged as the largest
population of identified TeV sources (see table~\ref{tab:gal}). As the
number of extended VHE $\gamma$-ray sources along the Galactic Plane
has increased, the likelihood of chance associations with pulsars is
now far from negligible. A systematic search for coincidences between
sources detected in the H.E.S.S. galactic plane survey and radio
pulsars has recently been performed by the H.E.S.S. collaboration
\cite{HESS:pwnpop_astroph}.  A clear excess of $\gamma$-ray nebulae in
positional coincidence with high spin-down luminosity pulsars (those
with $\dot{E}/d^{2}$ above $\sim$10$^{35}$ erg\,s$^{-1}$\,kpc$^{-2}$)
is found, in comparison to expectations for chance coincidences.  The
implied efficiency in the conversion of spin-down power, via
ultra-relativistic winds, into TeV $\gamma$-ray production is around
1\%.  A key recent result in this area, is that of energy-dependant
morphology in HESS\,J1825$-$137 \cite{HESS:1825p2}. New H.E.S.S. data
show that the $\gamma$-ray emission `shrinks' at high energies:
towards the pulsar PSR\,B1823$-$13. Such behaviour has been seen
before in X-ray synchrotron emission and has been interpreted as
evidence for the energy-losses of $>$ TeV electrons. The discovery of
this effect in $\gamma$-rays provides us with a potentially powerful
new tool with which to investigate high energy particles within these
objects.

The remaining well established class of galactic TeV sources is that of
\emph{binary systems} of a compact object and a massive star. Three such
systems have now been firmly identified (see table~\ref{tab:gal}). 
These objects appear to belong to one of two classes: microquasars or
binary PWN.  Whilst the 3.4~year period system of PSR B1259$-$63 and
the Be-star SS\,2883 certainly belongs to the later class, in the two
remaining well established systems, LS\,5039 and LS\,I\,+61\,303 the
acceleration site is not yet clear. In the binary pulsar scenario the
energy source is the spin-down of the neutron star, in the microquasar
scenario accretion is the power-source and the particle acceleration
occurs in relativistic jets produced close to the compact object
(black hole or neutron star). The best $\gamma$-ray microquasar
candidate so far is perhaps Cyg\,X-1, in which the compact object is
certainly a black hole. However, the evidence for TeV emission from
this object has not yet reached the level where a robust detection
claim can be made \cite{MAGIC:cygx1}.  See \cite{Mirabel:Binaries} for
a recent review of this topic.

Beyond these established TeV source classes there are indications of
an emerging class of sources associated with \emph{clusters of massive young
stars}. The colliding winds of massive stars are thought to result in
strong shocks capable of accelerating particles up to TeV energies
(see for example \cite{DomingoTorres:winds} and \cite{Pittard:WR140})
and particularly, the collective effect of such winds could be
detectable in $\gamma$-rays.  The recently discovered $\gamma$-ray
source HESS\,J1023$-$575 \cite{HESS:westerlund2} is coincident with
the massive stellar cluster Westerlund\,2: the second most massive
young cluster in our galaxy. Whilst this association may be
coincidental, the colliding winds of stars in this cluster can
certainly provide the energy required to produce the $\gamma$-ray
emission and acceleration in such objects seems plausible.

\subsection{Extragalactic Sources}

Extragalactic objects so far face none of the identification problems
of galactic sources, to date all are point-like and no serendipitous
discoveries have been made. Indeed, so far all extragalactic
$\gamma$-ray sources appear to be Active Galactic Nuclei (AGN).  AGN
are believed to host actively accreting $>10^{6}$ solar mass black
holes which drive powerful relativistic jets into their
environments.  The \emph{blazar} subclass of AGN is characterised 
rapid variability and broad-band non-thermal emission. These objects
are thought to represent AGN with jets aligned very closely
($<$10$^{\circ}$) with the line of sight to the observer, resulting in
fluxes enhanced through beaming effects.

Table~\ref{tab:agn} lists the known TeV emitting AGN in order of
redshift.  This table shows that more distant objects ($z>0.13$) have
been discovered only in the past two years. This is a consequence of
the attenuation of TeV photons via pair-production on the
extragalactic background light (EBL) (see for example
\cite{EBL:FazioStecker}). This absorption provides an effective
horizon to the universe which expands rapidly at low energies.
Sensitive instruments with low energy thresholds are therefore
required to detect distant objects. The distance corresponding at an
optical depth of one is approximately $z = 0.1$ at 1~TeV.  Only
relatively recently have experiments with substantial sensitivity in
the 0.05--1 TeV range existed, leading to a rapid expansion in the
number of $z>0.1$ TeV blazars. Table~\ref{tab:agn} reveals some
evidence for the expected softening of spectra at larger redshift, but
the intrinsic spread in blazar spectral properties is clearly very
large.

The distortion of $\gamma$-ray spectra by extragalactic attenuation,
though problematic for TeV studies of distant objects, can be used to
advantage in deriving limits on the wavelength dependent density of
the EBL.  Under the assumption that the intrinsic spectrum of these
objects has a photon index not less than 1.5 (that expected for
inverse Compton radiation of an $E^{-2}$ electron spectrum radiating
in the Thompson limit), limits on the near mid- and near infra-red EBL
have been calculated that approach the lower limits from galaxy counts
at these wavelengths, effectively resolving the EBL density at these
wavelengths \cite{HESS:ebl,HESS:1es0229,HESS:1es0347,EBL:MazinRaue}.
The spectrum of the EBL provides important constraints on the star 
formation history of the universe. For example, the TeV limits in the
near infra-red range can be used to place limits on the contribution of the
first generation of stars \cite{HESS:ebl}.   
The recent detection of the $z=0.536$ flat spectrum radio quasar 
3C\,279 by the MAGIC collaboration represents a major step forward 
in distance, and promises to be very important for constraining the 
EBL at shorter wavelengths~\cite{MAGIC:3c279icrc}.

\begin{table}
\footnotesize\rm
\caption{
The known very high energy $\gamma$-ray emitting AGN.
Only statistical errors are given on the photon index.
The final column gives the reference to the discovery
publication and also the reference for the photon index,
where different.
}
\begin{indented}
\item[]\begin{tabular}{|l|l|l|l|l|l|l|} \hline 
Object       &  Discovered &  Year & $z$   & Class & Photon Index &  Ref. \\\hline
M\,87         &  HEGRA      &  2003 & 0.004 & LINER &$2.22\pm0.15$& \cite{HEGRA:m87,HESS:m87} \\
Mrk\,421      &  Whipple    &  1992 & 0.031 & HBL   &$2.56\pm0.07$& \cite{Whipple:mrk421, Whipple:mrk421spec}     \\
Mrk\,501      &  Whipple    &  1996 & 0.034 & HBL   &$2.47\pm0.07$& \cite{Whipple:mrk501, HEGRA:mrk501spec}  \\
1ES\,2344+514 &  Whipple    &  1998 & 0.044 & HBL   &$2.54\pm0.17$& \cite{Whipple:1es2344, Whipple:1es2344spec}	      \\
Mrk\,180      &  MAGIC      &  2006 & 0.046 & HBL   &$3.3\pm0.7$  & \cite{MAGIC:mrk180}	      \\
1ES\,1959+650 &  TA         &  2002 & 0.047 & HBL   &$2.83\pm0.14$& \cite{TA:1es1959, HEGRA:1es1959}   \\
BL\,Lac       &  MAGIC      &  2006 & 0.069 & LBL   &$3.6\pm0.5$  & \cite{MAGIC:bllac_astroph}   \\
PKS\,0548-322 &  HESS       &  2006 & 0.069 & HBL   &     -       & \cite{HESS:pks0548_icrc} 	      \\
PKS\,2005-489 &  HESS       &  2005 & 0.071 & HBL   &$4.0\pm0.4$  &\cite{HESS:pks2005} 	      \\
PKS 2155-304 &  Durham     &  1999 & 0.116 & HBL   &$3.32\pm0.06$&\cite{Durham:pks2155,HESS:pks2155}\\
H\,1426+428   &  Whipple    &  2002 & 0.129 & HBL   &$3.50\pm0.35$&\cite{Whipple:h1426,Whipple:h1426spec}  \\
1ES\,0229+200 &  HESS       &  2007 & 0.140 & HBL   &$2.50\pm0.19$&\cite{HESS:1es0229} 	      \\
H\,2356-309   &  HESS       &  2005 & 0.165 & HBL   &$3.06\pm0.21$&\cite{HESS:ebl}	      \\
1ES\,1218+304 &  MAGIC      &  2005 & 0.182 & HBL   &$3.0\pm0.4$  &\cite{MAGIC:1es1218}	      \\
1ES\,1101-232 &  HESS       &  2005 & 0.186 & HBL   &$2.88\pm0.17$&\cite{HESS:ebl}	      \\
1ES\,0347-121 &  HESS       &  2007 & 0.188 & HBL   &$3.10\pm0.23$&\cite{HESS:1es0347} 	      \\
1ES\,1011+496 &  MAGIC      &  2007 & 0.212 & HBL   &$4.0\pm0.5$  &\cite{MAGIC:1es1011} \\
PG\,1553+113  &  HESS       &  2005 & $>0.25$ & HBL &$4.0\pm0.6$  & \cite{HESS:pg1553} \\
3C\,279       &  MAGIC      &  2007 & 0.536 & FSRQ  &    -        & \cite{MAGIC:3c279icrc}     \\
\hline
\end{tabular}
\end{indented}
\label{tab:agn}
\end{table}

\begin{figure}
\begin{center}
\includegraphics[width=0.9\textwidth]{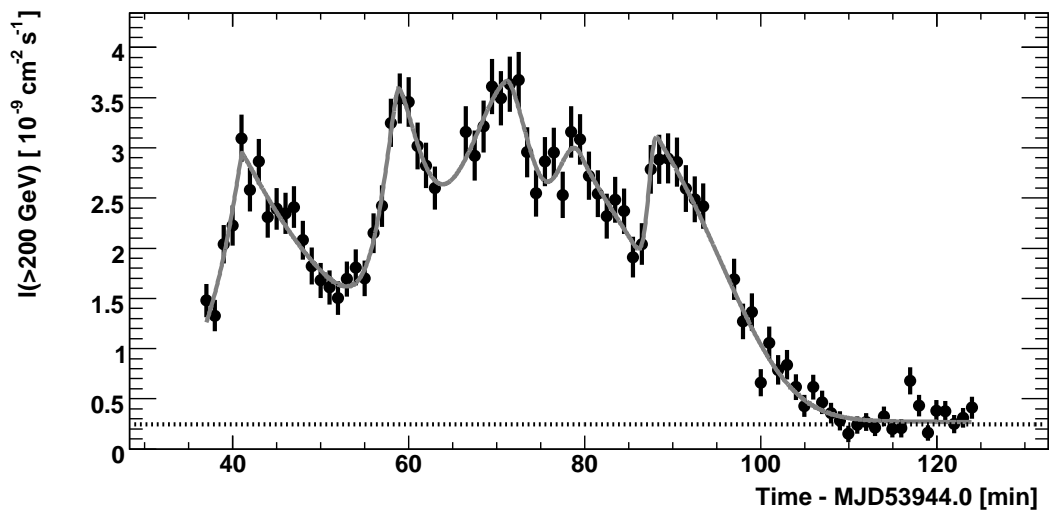}
\caption{$\gamma$-ray light curve, in one minute bins, of the spectacular flare from PKS\,2155$-$304 
  detected using H.E.S.S. in July 2006. Reproduced from ~\cite{HESS:pks2155_flare}.}
\label{fig:2155}
\end{center}
\end{figure}

The activity of the TeV blazar PKS\,2155$-$304 observed using
H.E.S.S. in July 2006 was the most dramatic seen so far from any
object in VHE $\gamma$-rays \cite{HESS:pks2155_flare}. Figure~\ref{fig:2155}
shows the light curve of the night with the highest flux, during which
the emission reached two orders of magnitude higher fluxes than those
typically seen from this object. Variability is clearly visible on
timescales of a few minutes in figure~\ref{fig:2155} and the best
measured individual flare is the first of the night with a best fit
rise-time of $173\pm23$ seconds.  Such rapid variability suggests an
extremely large Doppler factor ($\sim$100), approaching that seen for
Gamma-ray Bursts and extremely challenging for models. Activity from
Mrk 501 at a somewhat lower level was detected using MAGIC one year
earlier~\cite{MAGIC:mrk501astroph}, and is interesting due to the evidence
for \emph{time-lags} between activity in different VHE energy bands. 
The absence of lags has been used in the past to constrain any possible
energy dependence of the speed of light (due for example to quantum
gravity effects)~ \cite{Whipple:QG}. Whilst it is quite plausible that
the lags are intrinsic to the source due to the different 
acceleration and energy loss timescales of particles of different
energies, the MAGIC result may be the first hint for new physics
\cite{MAGIC:quantumgrav_astroph}. The H.S.S.S. data from 
PKS\,2155$-$304 should allow this hypothesis to be tested in the near
future.

\begin{figure}
\begin{center}
\includegraphics[width=0.7\textwidth]{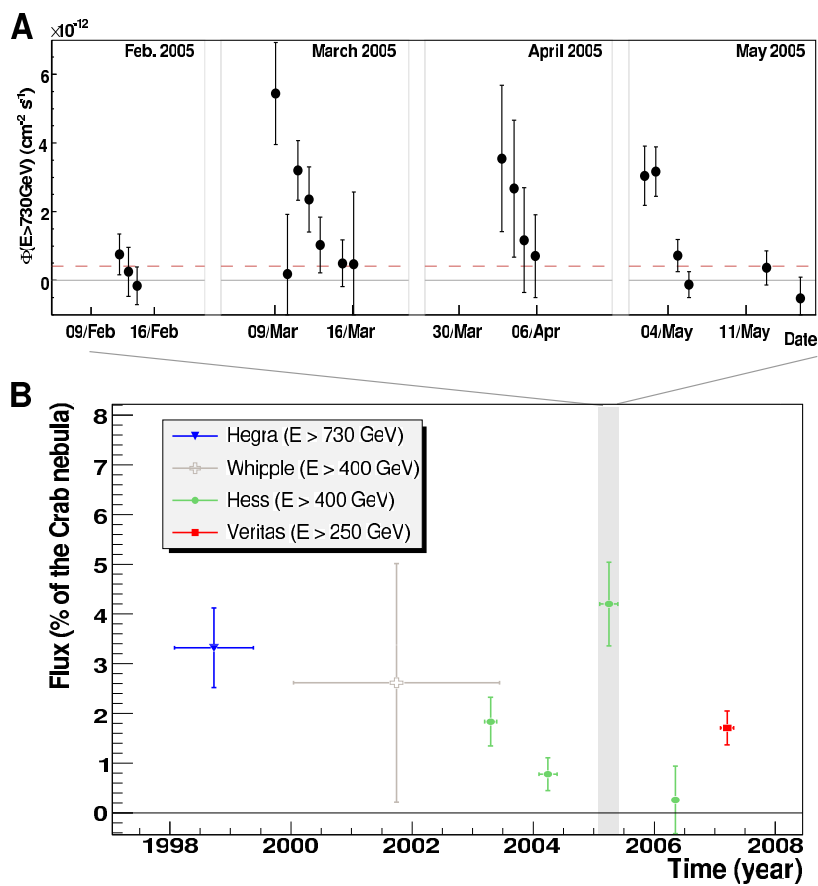}
\caption{Long and short-term variability in the TeV emission of M\,87.
A) Short-term variability seen in the light-curve of M\,87 using H.E.S.S. in 2005, reproduced from \cite{HESS:m87}
and B) Long-term variability as seen using HEGRA, Whipple, H.E.S.S. and VERITAS,
reproduced from \cite{VERITAS:m87_icrc}.
}
\label{fig:m87}
\end{center}
\end{figure}

The nearby radio galaxy M\,87 has a jet inclined at
$\sim$30$^{\circ}$ to the line of sight and is hence the only non-blazar extragalactic
TeV source. Given the reduced beaming effects in such a system, and the mass
of the black hole ($\approx$3$\times$10$^{9}$ solar masses), the two day timescale variability discovered 
using H.E.S.S. \cite{HESS:m87} is particularly surprising. Causality
arguments have been used to derive a limit of $5 \delta R_{s}$ on
the size of the emission region, where  $\delta$ is the Doppler factor
of the source and $R_{s}$ is the Schwarzschild radius of the 
supermassive black hole. Figure~\ref{fig:m87} shows the light-curve
of M\,87 on long (year) and short (day) timescales including data
from several VHE instruments. The most recent data shown are the
$5.1 \sigma$ detection of this source using VERITAS earlier 
this year \cite{VERITAS:m87_icrc}.

\section{Summary} 

Cherenkov telescope arrays have proved themselves to be the most
effective way to pursue (photon) astronomy at the highest energies.
Of the 71 currently known TeV $\gamma$-ray sources, 68 were discovered
using the imaging Cherenkov technique. Whilst the alternative
approach, based on shower particle detection at ground level, is
clearly complementary due to the much wider field of view and duty
cycle achievable, Cherenkov instruments are likely to remain the
work-horse of the field for some time.  The sensitivity and precision
obtain with this technique are unrivalled in the high energy ($>$ MeV)
domain. An order of magnitude improvement in sensitivity should be
achievable with next generation instruments such as CTA and AGIS,
which are discussed elsewhere in this issue.

The most important recent scientific highlights produced using this 
technique are perhaps the ongoing survey of our
galaxy with H.E.S.S., which has resulted in the discovery of a large 
fraction of the known TeV sources, and the recent detections of
distant ($z\gg0.1$) AGN using H.E.S.S. and MAGIC, marking very significant 
expansion of volume of the universe accessible to ground-based $\gamma$-ray
instruments. 
The potential for new discoveries seems to be very large, with
several source classes, for example Clusters of Galaxies and Starbursts,
predicted to emit at flux levels very close to current sensitivity limits.
VHE $\gamma$-ray astronomy using Cherenkov Telescope arrays can now be 
regarded as a well established astronomical discipline --- with a very 
bright future.

\section*{Acknowledgments}
I would like to thank all authors who agreed to provide plots and
J. Skilton for her careful reading of the manuscript.


\section*{References}

\begin{thebibliography}{108}

\bibitem{Whipple:crab}
T.~C. {Weekes} et al.
 1989 \newblock {\em \apj}  342, 379 

\bibitem{Hofmann:Performance}
W.~{Hofmann}
2006 \newblock {\em ArXiv e-prints} astro-ph/0603076

\bibitem{MAGIC:moon}
J.~{Rico}, E~{de Ona-Wilhelmi}, J.~{Cortina}, and E.~{Lorenz}.
 2007 \newblock {\em ArXiv e-prints}  0709.2283  

\bibitem{GC:Whipple04}
K.~{Kosack} et al.
 2004 \newblock {\em \apjl}  608, L97 

\bibitem{HESS:mrk421}
F.~{Aharonian} et al.
 2005 \newblock {\em \aap}  437, 95 

\bibitem{5at5}
F.~A. {Aharonian}, A.~K. {Konopelko}, H.~J. {V{\"o}lk}, and H.~{Quintana}.
 2001 \newblock {\em Astroparticle Physics}  15, 335 

\bibitem{Biller:WideFoV}
I.~{de La Calle P{\'e}rez} and S.~D. {Biller}.
 2006 \newblock {\em Astroparticle Physics}  26, 69 

\bibitem{hillas85:technique}
A.~M. {Hillas}
1985 \newblock{\em 19th International Cosmic Ray Conference, La Jolla, USA,} 3, 445

\bibitem{hillas96:technique}
A.~M. {Hillas}
1996 \newblock{\em Space Science Reviews} 75, 17

\bibitem{CAT:model}
F. Piron et al.
2001  \newblock {\em \aap} 374, 895

\bibitem{deNaurois:Model}
M.~{de Naurois}
2006 \newblock {\em ArXiv e-prints} astro-ph/0607247 

\bibitem{HESS:model_marianne}
M.~{Lemoine-Goumard}, B.~{Degrange}, and M.~{Tluczykont}
2006 \newblock {\em Astroparticle Physics}  25, 195 

\bibitem{Berge:background}
D.~{Berge}, S.~{Funk}, and J.~{Hinton}
 2007 \newblock {\em \aap}  466, 1219 

\bibitem{HEGRA:performance1}
A.~{Daum} et al.
 1997 \newblock {\em Astroparticle Physics}  8, 1 

\bibitem{HESS:trigger}
S.~{Funk} et al.
2004 \newblock {\em Astroparticle Physics} 22, 285

\bibitem{mono_rec}
R. W. Lessard et al.
2001 \newblock {\em Astroparticle Physics}  15, 1 

\bibitem{HEGRA:performance2}
A.~{Konopelko} et al.
 1999 \newblock {\em Astroparticle Physics}  10, 275 

\bibitem{Mirzoyan:Muons}
R.~{Mirzoyan}, D.~{Sobczynska}, E.~{Lorenz}, and M.~{Teshima}
 2006 \newblock {\em Astroparticle Physics}  25, 342 

\bibitem{Cherenkov:Timing}
V.~R. {Chitnis} and P.~N. {Bhat}
 2001 \newblock {\em Astroparticle Physics}  15, 29 

\bibitem{MAGIC:timing}
D.~{Tescaro} et al.
 2007 \newblock {\em ArXiv e-prints}  0709.1410

\bibitem{HESS:status}
J.~A. {Hinton}
 2004 \newblock {\em New Astronomy Review}  48, 331 

\bibitem{HESS:optics}
K.~{Bernl{\"o}hr} et al.
2003 \newblock {\em Astroparticle Physics} 20, 111

\bibitem{HESS:phaseII}
P. Vincent et al
2005 \newblock {\em 29th International Cosmic Ray Conference, Pune, India} 5, 163  

\bibitem{MAGIC:status}
J.~{Cortina}
 2005 \newblock {\em \apss}  297, 245 

\bibitem{MAGIC:technical}
J.~{Cortina} et al.
 2005 \newblock  {\em 29th International Cosmic Ray Conference, Pune, India} 5, 359  

\bibitem{MAGIC-II:camera}
C.~C. {Hsu} et al.
 2007 \newblock {\em ArXiv e-prints} 0709.2474,  

\bibitem{VERITAS:first_tel}
J.~{Holder} et al.
 2006 \newblock {\em Astroparticle Physics}  25, 391 

\bibitem{VERITAS:status}
J.~{Holder} et al.
2006 \newblock {\em ArXiv e-prints} astro-ph/0611598

\bibitem{VERITAS:icrc}
G.~{Maier} et al.
 2007 \newblock {\em ArXiv e-prints}  0709.3654,  

\bibitem{CANGAROO3:status}
H.~{Kubo} et al.
 2004 \newblock {\em New Astronomy Review}  48, 323 

\bibitem{CANGAROO:3.8m}
S.~{Ebisuzaki} et al.
1991 \newblock {\em 22nd International Cosmic Ray Conference, Dublin, Ireland} 2, 607  

\bibitem{CANGAROO2:status}
T.~{Yoshikoshi} et al.
 1999 \newblock {\em Astroparticle Physics}  11, 267 

\bibitem{CANGAROO:icrc}
M. Mori et al.
2007 \newblock {\em 30th International Cosmic Ray Conference, Merida, Mexico}

\bibitem{TACTIC:NIM}
R.~{Koul} et al.
2007 \newblock {\em Nuc. Instr. and Meth. in Phys. Research} A 578, 548

\bibitem{PACT:agn}
D.~{Bose} et al.
 2007 \newblock {\em \apss}  309, 111 

\bibitem{CELESTE:status}
J.~{Bussons Gordo}
 2004 \newblock {\em New Astronomy Reviews}  48, 351 

\bibitem{STACEE:status}
D.~A. {Williams} et al.
 2004 \newblock {\em New Astronomy Reviews}  48, 359 

\bibitem{SolarTwo:status}
T.~{T{\"u}mer} et al.
 1999 \newblock {\em Astroparticle Physics}  11, 271 

\bibitem{GRAAL:crab}
F.~{Arqueros} et al.
 2002 \newblock {\em Astroparticle Physics}  17, 293 

\bibitem{HESS:scanpaper1}
F.~{Aharonian} et al.
2005 \newblock {\em Science} 307, 1938

\bibitem{HESS:scanpaper2}
F.~{Aharonian} et al.
2006 \newblock {\em \apj} 636, 777

\bibitem{HESS:scanicrc}
S.~Hoppe et al.
2007 \newblock {\em ArXiv e-prints} 0710.3528

\bibitem{HESS:psrb1259}
F.~{Aharonian} et al.
 2005 \newblock {\em \aap}  442, 1 

\bibitem{HESS:ls5039p1}
F.~{Aharonian} et al.
 2005 \newblock {\em Science}  309, 746 

\bibitem{HESS:ls5039p2}
F.~{Aharonian} et al.
 2006 \newblock {\em \aap}  460, 743 

\bibitem{MAGIC:lsi61}
J.~{Albert} et al.
 2006 \newblock {\em Science}  312, 1771 

\bibitem{VERITAS:lsi61}
V.~A. {Acciari} et al.
 2008 \newblock {\em ArXiv e-prints} 0802.2363,  

\bibitem{CANGAROO:rxj1713}
H.~{Muraishi} et al.
 2000 \newblock {\em \aap}  354, L57 

\bibitem{HESS:rxj1713p1}
F.~A. {Aharonian} et al.
 2004 \newblock {\em \nat}  432, 75 

\bibitem{HESS:rxj1713p2}
F.~{Aharonian} et al.
 2005 \newblock {\em \aap}  449, 223 

\bibitem{CANGAROO:velajnr}
H. Katagiri et al.
2005 \newblock {\em \apjl} 619, L163

\bibitem{HESS:velajnr}
F.~{Aharonian} et al.
 2005 \newblock {\em \aap}  437, L7 

\bibitem{HESS:velajnr2}
F.~{Aharonian} et al.
 2007 \newblock {\em \apj}  661, 236 

\bibitem{HESS:rcw86_icrc}
S. Hoppe et al.
2007 \newblock {\em 30th International Cosmic Ray Conference, Merida, Mexico}

\bibitem{HEGRA:casA}
F.~{Aharonian} et al.
 2001 \newblock {\em \aap}  370, 112 

\bibitem{Whipple:crabspec}
A.~M. {Hillas} et al.
 1998 \newblock {\em \apj}  503, 744 

\bibitem{HESS:msh1552}
F.~{Aharonian} et al.
 2005 \newblock {\em \aap}  435, L17 

\bibitem{HESS:velax}
F.~{Aharonian} et al.
 2006 \newblock {\em \aap}  448, L43 

\bibitem{HESS:1825p1}
F.~A. {Aharonian} et al.
 2005 \newblock {\em \aap}  442, L25 

\bibitem{HESS:1825p2}
F.~{Aharonian} et al.
 2006 \newblock {\em \aap}  460, 365 

\bibitem{HESS:kookaburra}
F.~{Aharonian} et al.
 2006 \newblock {\em \aap}  456, 245 

\bibitem{HESS:g09}
F.~{Aharonian} et al.
 2005 \newblock {\em \aap}  432, L25 

\bibitem{HESS:rxj1713p3}
F.~{Aharonian} et al.
 2007 \newblock {\em \aap}  464, 235 

\bibitem{Berezhko:rxj1713}
E.~G. {Berezhko} and H.~J. {V{\"o}lk}
 2006 \newblock {\em \aap}  451, 981 

\bibitem{Kelner:pp}
S.~R. {Kelner}, F.~A. {Aharonian}, and V.~V. {Bugayov}
 2006 \newblock {\em \prd}  74, 034018 

\bibitem{Porter:ic_snr}
T.~A. {Porter}, I.~V. {Moskalenko}, and A.~W. {Strong}
 2006 \newblock {\em \apjl}  648, L29 

\bibitem{HESS:w28_icrc}
{G.~Rowell} et al.
 2007 \newblock {\em ArXiv e-prints}  0710.2017

\bibitem{MAGIC:ic443}
J.~{Albert} et al.
 2007 \newblock {\em \apjl}  664, L87 

\bibitem{VERITAS:ic443_icrc}
T. B. Humensky et al.
2007 \newblock {\em 30th International Cosmic Ray Conference, Merida, Mexico}

\bibitem{HESS:gc_diffuse}
F.~{Aharonian} et al.
 2007 \newblock {\em Nature} 439, 695 

\bibitem{HESS:pwnpop_astroph}
S.~{Carrigan} et al.
2007 \newblock {\em ArXiv e-prints} 0709.4094

\bibitem{MAGIC:cygx1}
J.~{Albert} et al.
 2007 \newblock {\em \apjl}  665, L51 

\bibitem{Mirabel:Binaries}
I.~F. {Mirabel}
 2006 \newblock {\em Science}  312, 1759 

\bibitem{DomingoTorres:winds}
E.~{Domingo-Santamar{\'{\i}}a} and D.~F. {Torres}
 2006 \newblock {\em \aap}  448, 613 

\bibitem{Pittard:WR140}
J.~M. {Pittard} and S.~M. {Dougherty}
 2006 \newblock {\em \mnras}  372, 801 

\bibitem{HESS:westerlund2}
F.~{Aharonian} et al.
 2007 \newblock {\em \aap}  467, 1075 

\bibitem{EBL:FazioStecker}
G.~G. {Fazio} and F.~W. {Stecker}
 1970 \newblock {\em \nat}  226, 135 

\bibitem{HESS:ebl}
F.~{Aharonian} et al.
 2006 \newblock {\em \nat}  440, 1018 

\bibitem{HESS:1es0229}
F.~{Aharonian} et al.
 2007 \newblock {\em \aap} 475, L9

\bibitem{HESS:1es0347}
F.~{Aharonian} et al.
 2007 \newblock {\em \aap}  473, L25 

\bibitem{EBL:MazinRaue}
D.~{Mazin} and M.~{Raue}
 2007 \newblock {\em \aap}  471, 439 

\bibitem{MAGIC:3c279icrc}
M.~{Teshima} et al.
 2007 \newblock {\em ArXiv e-prints}  0709.1475,  

\bibitem{HEGRA:m87}
F.~{Aharonian} et al.
 2003 \newblock {\em \aap}  403, L1 

\bibitem{HESS:m87}
F.~{Aharonian} et al.
 2006 \newblock {\em Science}  314, 1424 

\bibitem{Whipple:mrk421}
M.~{Punch} et al.
 1992 \newblock {\em \nat}  358, 477 

\bibitem{Whipple:mrk421spec}
J.~A. {Zweerink} et al.
 1997 \newblock {\em \apjl}  490, L141

\bibitem{Whipple:mrk501}
J.~{Quinn} et al.
 1996 \newblock {\em \apjl}  456, L83 

\bibitem{HEGRA:mrk501spec}
F.~{Aharonian} et al.
 1997 \newblock {\em \aap}  327, L5 

\bibitem{Whipple:1es2344}
M.~{Catanese} et al.
 1998 \newblock {\em \apj}  501, 616 

\bibitem{Whipple:1es2344spec}
M.~{Schroedter} et al.
 2005 \newblock {\em \apj}  634, 947 

\bibitem{MAGIC:mrk180}
J.~{Albert} et al.
 2006 \newblock {\em \apjl}  648, L105 

\bibitem{TA:1es1959}
T. Nishiyama et al.
 1999 \newblock {\em 26th International Cosmic Ray Conference, Salt Lake City, USA} 3, 370  

\bibitem{HEGRA:1es1959}
F.~{Aharonian} et al.
 2003 \newblock {\em \aap}  406, L9 

\bibitem{MAGIC:bllac_astroph}
J. Albert et al
2007 \newblock {\em \apjl} 666, L17

\bibitem{HESS:pks0548_icrc}
G. Superina, et al.
2007 \newblock {\em 30th International Cosmic Ray Conference, Merida, Mexico}

\bibitem{HESS:pks2005}
F.~{Aharonian} et al.
 2005 \newblock {\em \aap}  436, L17 

\bibitem{Durham:pks2155}
P.~M. {Chadwick} et al.
 1999 \newblock {\em \apj}  513, 161 

\bibitem{HESS:pks2155}
F.~{Aharonian} et al.
 2005 \newblock {\em \aap}  430, 865 

\bibitem{Whipple:h1426}
D.~{Horan} et al.
 2002 \newblock {\em \apj}  571, 753 

\bibitem{Whipple:h1426spec}
D.~{Petry} et al.
 2002 \newblock {\em \apj}  580, 104 

\bibitem{MAGIC:1es1218}
J.~{Albert} et al.
 2006 \newblock {\em \apjl}  642, L119 

\bibitem{MAGIC:1es1011}
J.~{Albert} et al.
 2007 \newblock {\em \apjl}  667, L21 

\bibitem{HESS:pg1553}
F.~{Aharonian} et al.
 2006 \newblock {\em \aap}  448, L19 

\bibitem{HESS:pks2155_flare}
F.~{Aharonian} et al.
 2007 \newblock {\em \apjl}  664, L71 

\bibitem{MAGIC:mrk501astroph}
J. Albert et al.	
 2007 \newblock {\em \apj}  669, 862 

\bibitem{Whipple:QG}
S.~D. {Biller} et al.
 1999 \newblock {\em Physical Review Letters}  83, 2108 

\bibitem{MAGIC:quantumgrav_astroph}
J.~{Albert} et al.
 2007 \newblock {\em ArXiv e-prints} 0708.2889

\bibitem{VERITAS:m87_icrc}
P. Colin et al.
 2007 \newblock {\em 30th International Cosmic Ray Conference, Merida, Mexico}

\end{thebibliography}
\end{document}